\documentclass{article}

\usepackage[T2A]{fontenc}
\usepackage[utf8]{inputenc}

\usepackage[english]{babel}

\usepackage[numbers,sort&compress]{natbib}

\usepackage{amsmath,amssymb}
\usepackage{graphicx}
\usepackage[dvipsnames]{xcolor}
\usepackage[export]{adjustbox}
\usepackage{url}
\usepackage{multido}
\usepackage{geometry}
\usepackage[linktocpage,unicode,colorlinks=true,linkcolor=black,citecolor=blue,urlcolor=blue]{hyperref}
\DeclareUnicodeCharacter{0096}{--}

\begin{document}
	
	\title{Continuum Landscape of stable extra-dimensional metrics}
\author{{Sergey~G.~Rubin}$^{a}$\thanks{e-mail: sergeirubin@list.ru}}

\date{\em \small
	$^a$ National Research Nuclear University MEPhI \\
	 (Moscow Engineering Physics Institute), \\
	 115409, Kashirskoe shosse 31, Moscow, Russia}


\maketitle

\begin{abstract}
	
	We investigate modified gravity with quadratic curvature terms
	($R^2$, $R_{AB}R^{AB}$, and $R_{ABCD}R^{ABCD}$). In this framework, 	the geometry of the extra dimensions is determined by the higher-dimensional field equations together with auxiliary conditions. For inhomogeneous metrics, we derive new stability conditions under radion excitations. The central result is that stable solutions satisfying these conditions are not isolated: a typical stable regular metric belongs to a continuous family of stable metrics in its immediate
	neighbourhood. Since the effective four-dimensional physical parameters depend on the extra-dimensional geometry, this geometrical landscape of stable metrics induces a continuous landscape of effective 	four-dimensional parameters. Transitions between distinct static solutions within the landscape cannot be induced by spatially localized
	perturbations, whose asymptotic decay ($|x|\to\infty$) preserves the boundary data distinguishing the states. We also note a possible topological transition from a two-brane to a one-brane configuration, associated with a change in the number of regular zeros of the metric	function \(r(u)\).
	
\end{abstract}

\section{Introduction}

The idea that  four-dimensional physical parameters may emerge from a higher-dimensional theory is one of the central themes of modern theories with extra dimensions. A prominent realization of this idea is the string landscape~\cite{Susskind:2003kw}, in which different effective four-dimensional theories arise from distinct higher-dimensional vacuum configurations. Here we explore a complementary possibility: a continuum of classical metric admitted by a fixed higher-dimensional theory may give rise to a landscape~\cite{2002PhRvD..66b4036C,Petriakova:2023klf,Rubin:2015pqa}.

We define the geometrical landscape as the set of admissible auxiliary conditions \(p\) that generate distinct classical extra-dimensional geometries \(g_n(u;p)\) for fixed parameters of the higher-dimensional action. Here \(p\) denotes a set of the initial or boundary conditions required to solve the equations of motion. Each set \(p\) corresponds to a particular extra-dimensional geometry and, after dimensional reduction, to a specific realization of the effective four-dimensional theory:
\begin{equation}\label{la4D}
	S_{4+n}(\lambda_{4+n},g_n) \longrightarrow S_4(\lambda_4[g_n]),
\end{equation}
where the effective four-dimensional parameters \(\lambda_4\) are functionals of the extra-dimensional geometry \(g_n\).

 
In this work we focus on the local stability with respect to a
radion-like breathing mode and derive the conditions for the absence of ghost and tachyonic instabilities. Our central result is that these stability conditions constrain the family of classical extra-dimensional solutions without, in general, reducing it to a single configuration: a continuous family of stable solutions survives after the physical requirement of radion stability is imposed. Thus, the resulting landscape of effective four-dimensional parameters reflects the continuous degeneracy of the classical extra-dimensional solutions, while the fundamental parameters \(\lambda_{4+n}\) of the higher-dimensional
action remain fixed.

The paper is organized as follows. In Section 2 we present the higher-dimensional action and the metric ansatz. Section 3 is devoted to the dimensional reduction to four dimensions and the derivation of the effective radion action in both the Jordan and Einstein frames. In Section 4 we formulate the conditions for the absence of ghosts and tachyonic instabilities of the radion. Section 5 contains the central result: under the stated smoothness and non-degeneracy assumptions, typical stable solution belongs to a continuous family of stable solutions. Section 6 discusses a topological transition between one-brane and two-brane configurations. Finally, Section 7 summarizes our findings and outlines open questions.

\section{The choice of Lagrangian}

We consider a \(D\)-dimensional gravitational theory with quadratic curvature corrections. A general \(D\)-dimensional quadratic-curvature gravity action that includes an arbitrary \(f(R)\) term is taken in the form
\begin{equation}\label{S}
	S=\frac12 m_D^{D-2}\int d^D x \sqrt{|g_D|}
	\Big[f(R)+c_1R_{AB}R^{AB}
	+c_2R_{ABCD}R^{ABCD}\Big].
\end{equation}
This form provides a convenient basis for a higher-dimensional effective theory~\cite{2007ARNPS..57..329B} containing the function of the Ricci scalar \(f(R)\) and quadratic curvature invariants --- the Ricci tensor squared \(R_{AB}R^{AB}\) and the Kretschmann scalar  \(R_{ABCD}R^{ABCD}\). Here \(R\) is the \(D\)-dimensional Ricci scalar, \(m_D\) is the \(D\)-dimensional Planck mass, and \(c_1\) and \(c_2\) are the parameters of the quadratic curvature corrections. 

The generalized Einstein equations
\begin{equation}\label{EE}
	\begin{aligned}
		& f'(R)R_{MN} - \frac{1}{2}f(R)g_{MN} + \left(\nabla_M\nabla_N - g_{MN}\Box\right)f'(R) \\
		& \quad + c_1 \left( 2R_{MA}R^A_{\ N} - \frac{1}{2}g_{MN}R_{AB}R^{AB} + \nabla_A\nabla_B\left(2\delta^A_{(M}R^B_{\ N)} - g_{MN}R^{AB}\right) \right) \\
		& \quad + c_2 \left( 2R_{MABC}R_N^{\ ABC} - \frac{1}{2}g_{MN}R_{ABCD}R^{ABCD} + 4\nabla^A\nabla^B R_{MANB} \right) = 0.
	\end{aligned}
\end{equation}
can be resolved in some particular  cases. Here we restrict the function \(f(R)\) to the quadratic form
\begin{equation}
	f(R)=aR^2+bR+c,
\end{equation}
where, without loss of generality, \(b=\pm 1\). Here we choose \(b=1\) and \(a>0\). The application of the slow motion approximation \cite{Shafi:1986vv,Bronnikov:2005iz} considerably simplifies the analysis.

We consider a spacetime with \(D=4+n\) dimensions and assume that the metric of the extra-dimensional space is static and depends on a single coordinate \(u\). Specifically, the static inhomogeneous metric has the form
\begin{equation}\label{met}
	ds^2=e^{2\gamma(u)}g_{\mu\nu}dx^\mu dx^\nu
	-\left[du^2+r^2(u)d\Omega_{n-1}^2\right].
\end{equation}

Hereafter, the warp factor is set to $\gamma (u) = 0$, unless stated otherwise. Here \(g_{\mu\nu}\) is the four-dimensional metric close to the Minkowski one, and \(d\Omega_{n-1}^2\) is the metric on a unit \((n-1)\)-dimensional sphere. Inhomogeneous extra dimensions of this type have been discussed, for example, in~\cite{Shiraishi1985,HuShen1986,ShenSobczyk1987,Dzhunushaliev:2008gf}.

The function \(r(u)\) determines the geometry of the extra dimensions and is assumed to be a solution to Eqs.~\eqref{EE}. We discuss a particular solution at the end of Section~4.

\section{Reduction to 4D}

\subsection{Choice of metric and auxiliary conditions}

We now derive the four-dimensional effective theory that governs the dynamics of a radion-like deformation of a given extra-dimensional geometry, following the standard dimensional-reduction procedure in the slow-motion (adiabatic) approximation; see, e.g., Refs.~\cite{Das:2017htt,Bhattacharyya:2025xws}. The background configuration in the extra dimensions is assumed to satisfy the static higher-dimensional field equations, while its size is allowed to vary slowly with the four-dimensional coordinates. Moduli field stabilization by bulk scalar fields has been investigated in higher-dimensional brane models~\cite{FlachiGarrigaPujolasTanaka2003}.
More specifically, we assume that a static solution \(r(u)\) is known and consider a perturbation of the internal metric of the form
\begin{equation}\label{metbet}
	ds^2=g_{\mu\nu}dx^\mu dx^\nu
	-e^{2\beta(x)}
	\left[du^2+r(u;p)^2d\Omega_{n-1}^2\right],
\end{equation}
where \(\beta(x)\) is the radion-like field. We retain terms up to second order in the four-dimensional derivatives of \(\beta\), while keeping the dependence on the extra-dimensional coordinate \(u\) exact. The slow variation of the metric is governed by the slow dynamics of the field \(\beta(x)\) over the spacetime coordinates. 

The metric function $r(u;p)$ belongs to the family of solutions to the static generalized Einstein equations \eqref{EE} in the Minkowski metric. Assuming the smoothness of the solutions, $r(u)$ is in one-to-one correspondence with the set of initial condition parameters $p$, defined as
\begin{equation}\label{param}
\{p\}=	\{p_1 = r(0),\quad p_2 = r'(0),\quad p_3 = R'(0)\}.
\end{equation}

In the \(f(R)\) framework the Ricci scalar \(R(u)\) propagates as an independent dynamical variable, so the system consists of two second-order equations, one for \(r(u)\) and one for \(R(u)\). A general solution is therefore specified by four initial data, \(r(0)\), \(r'(0)\), \(R(0)\), and \(R'(0)\). The field equations also contain a first integral, which imposes one constraint on these data, reducing the number of independent auxiliary parameters to three. 
The first-order constraint is used to fix one of the initial conditions — we choose  \(R(0)\) — and serves as a check of the numerical accuracy during the integration.

\subsection{Slow-motion approximation}

For metric~\eqref{metbet}, the \(D\)-dimensional Ricci scalar reads
\begin{equation}\label{Ricci_total}
	R=	R_4	+e^{-2\beta} \widetilde R_n(u)
	+2n\Box_4\beta	+n(n+1)g^{\mu\nu}
	\partial_\mu\beta\,\partial_\nu\beta ,
\end{equation}
in the notation of Birrell and Davies~\cite{Birrell:1982ix} (used throughout).  The quantity \(\widetilde R_n(u)\) is the scalar curvature of the unperturbed internal metric,
\begin{equation}
	\widetilde R_n(u)
	=(n-1)\left[
	(n-2)\frac{1-r'^2}{r^2}
	-2\frac{r''}{r}
	\right].
\end{equation}
Thus,
\begin{equation}
	R=R_4+R_n+P_k,	\qquad
	R_n=e^{-2\beta}\widetilde R_n,
\end{equation}
with
\begin{equation}
	P_k=	n(n+1)(\partial\beta)^2+2n\Box_4\beta,\quad (\partial\beta)^2\equiv(\partial_\mu\beta \partial^\mu\beta)
\end{equation}

The slow-motion approximation corresponds to
\begin{equation}\label{small}
	|R_4+P_k|\ll |R_n|,
\end{equation}
so that the four-dimensional curvature and the radion-derivative contributions can be treated as perturbations of the internal curvature; see the discussion in~\cite{Bronnikov:2005iz}. It it is assumed that each derivative $\partial_\mu$  contains a small parameter $\varepsilon$, so that $\partial_\mu \to \varepsilon \partial_\mu$ and $R_4 = \mathcal{O}(\varepsilon^2)$.

The curvature-squared invariants that enter the higher-dimensional action must be expanded consistently to the same order in four-dimensional derivatives. One finds

\begin{equation}
	\begin{aligned}
		R_{AB}R^{AB} ={}& e^{-4\beta}\left[
		(n-1)^2\left(\frac{r''}{r}\right)^2
		+(n-1)\left((n-2)\frac{1-r'^2}{r^2}-\frac{r''}{r}\right)^2 \right] \\
		&+2\widetilde R_n\, e^{-2\beta}\left[\Box\beta+ n(\partial\beta)^2\right]		+O(\epsilon^4),
	\end{aligned}
\end{equation}
and
\begin{equation}
	\begin{aligned}
		R_{ABCD}R^{ABCD} ={}& e^{-4\beta}\left[
		4(n-1)\left(\frac{r''}{r}\right)^2
		+2(n-1)(n-2)\frac{(1-r'^2)^2}{r^4}
		\right] \\
		&+4\widetilde R_n\, e^{-2\beta}(\partial\beta)^2
		+O(\epsilon^4).
	\end{aligned}
\end{equation}

For future needs we introduce the purely internal curvature combinations
\begin{equation}
	\begin{aligned}
		Q_1(r(u))=
		(n-1)^2\left(\frac{r''}{r}\right)^2
		+(n-1)\left[
		(n-2)\frac{1-r'^2}{r^2}
		-\frac{r''}{r}
		\right]^2 ,
	\end{aligned}
\end{equation}
and
\begin{equation}
	Q_2(r(u))=
	4(n-1)\left(\frac{r''}{r}\right)^2
	+2(n-1)(n-2)
	\frac{(1-r'^2)^2}{r^4}.
\end{equation}

\subsection{The Jordan frame}

Within the same approximation the function \(f(R)\) can be expanded as
\begin{equation}
	f(R)=
	f(R_n)+f_R(R_n)\left(R_4+P_k\right)
	+O(\epsilon^4).
\end{equation}
Substituting these expressions into the \(D\)-dimensional action and integrating over the internal coordinates, we obtain the effective four-dimensional action
\begin{equation}\label{S4}
	S=\frac12m_D^{D-2}
	\int d^4x\sqrt{|g_4|}
	\left[
	F(\beta)R_4
	+K(\beta)(\partial\beta)^2
	-2V(\beta)
	\right].
\end{equation}
Here \(v_{n-1}\) denotes the volume of the unit \((n-1)\)-sphere, and
\begin{equation}\label{FKV}
	F(\beta)=
	v_{n-1}\int du\,r^{n-1}
	e^{n\beta}
	f_R\left(\widetilde R_n e^{-2\beta}\right),
\end{equation}
while the coefficient of the radion kinetic term is
\begin{equation}\label{KJ}
	K(\beta)= 	n(n+1)F(\beta) 	-2nF'(\beta) 	+4(c_1+c_2)G(\beta),
\end{equation}
where the prime on \(F\) denotes differentiation with respect to \(\beta\) and
\begin{equation}\label{GKV}
	G(\beta)= 	v_{n-1}\int du\,r^{n-1}
	\widetilde R_n e^{(n-2)\beta}.
\end{equation}
 The second term in \eqref{KJ} arises from integration by parts of the \(\Box_4\beta\) contribution in the Ricci scalar.

Finally, the effective potential reads
\begin{equation}\label{VJ}
	V(\beta)=
	-\frac{v_{n-1}}{2}	\int du\,r^{n-1}
	\left\{	e^{n\beta}	f\left(\widetilde R_n e^{-2\beta}\right)
	+	e^{(n-4)\beta}	\left[c_1Q_1(u)+c_2Q_2(u)\right]
	\right\}.
\end{equation}
Hereafter, the limits of integration are defined by the zeros of the metric function \(r(u;p)\).

Thus the dimensional reduction maps a given classical extra-dimensional solution \(r(u;p)\) into a four-dimensional scalar-tensor theory whose kinetic coefficient and effective potential depend explicitly on the geometry of the extra dimensions. This dependence is essential for the stability analysis developed below.

\subsection{The Einstein frame}

Having derived the reduced four-dimensional action in the Jordan (physical) frame, it is convenient for the subsequent stability analysis to pass to the Einstein frame, where the scalar-tensor structure takes its canonical form and the gravitational sector decouples from the radion at the level of the kinetic term. We introduce the Einstein-frame metric by the conformal transformation
\begin{equation}
	g_{\mu\nu}^{(E)}=F(\beta)g_{\mu\nu}.
\end{equation}
The four-dimensional Ricci scalar transforms accordingly. After discarding a total-derivative term, the action becomes
\begin{equation}\label{S4E}
	S=\frac12m_D^{D-2}
	\int d^4x\sqrt{|g_4^{(E)}|}	\left[	R_4^{(E)}
	+K_{\rm E}(\beta)(\partial\beta)^2 -2V_{\rm E}(\beta)	\right],
\end{equation}
where
\begin{equation}\label{KEin}
	K_{\rm E}(\beta)	= 	\frac{K(\beta)}{F(\beta)}
	+\frac32	\left(\frac{F'(\beta)}{F(\beta)}\right)^2
\end{equation}
and
\begin{equation}\label{VEin}
	V_{\rm E}(\beta)	=	\frac{V(\beta)}{F^2(\beta)}.
\end{equation}
Equations~\eqref{KEin} and~\eqref{VEin} show explicitly how the original kinetic function \(K(\beta)\) and potential \(V(\beta)\) of the Jordan frame are transformed by the conformal factor \(F(\beta)\): the potential is simply rescaled by \(F^{-2}\), while the kinetic term acquires an additional geometric contribution \(\tfrac32(F'/F)^2\) that is generated by the conformal transformation itself and is present even when \(K(\beta)=0\).

\section{Radion stability}

\subsection{Stability conditions for inhomogeneous metric}\label{StabInhomo}

The existence of a physically acceptable ground state of the theory --- in particular, one that reproduces ordinary four-dimensional gravity together with a stable value of the internal-space size --- imposes a set of conditions on the Einstein-frame potential \(V_E\) and kinetic function \(K_E\).

We emphasize that our results should be understood as establishing the existence of a continuous family of configurations that are stable at least with respect to the radion degree of freedom. This constitutes a necessary condition for physical viability and represents the standard first step in the study of higher-dimensional gravitational models of this type.

First, we consider the stability of the metric~\eqref{met}. The point \(\beta=0\) must be an extremum of the effective potential:
\begin{equation}\label{extr}
	V_E'(\beta=0)=0.
\end{equation}
This condition is necessary for the stability of the background metric~\eqref{met} (where \(\beta=0\)).

Second, the absence of ghost degrees of freedom in the effective theory requires the kinetic function to be strictly positive at the extremum,
\begin{equation}\label{Kpos}
	K_E(\beta=0)=\Delta_K>0,
\end{equation}
which guarantees that the radion field has the correct sign of the kinetic energy and does not propagate as a ghost.

Third, the absence of a tachyonic instability of the radion around \(\beta=0\) requires the extremum to be a local minimum of the potential,
\begin{equation}\label{Vpppos}
	V_E''(\beta=0)=\Delta_V>0,
\end{equation}
which ensures that small perturbations of the internal-space size oscillate or decay rather than grow, so that the configuration is dynamically stable.

The stability conditions can be expressed directly in terms of the
quantities defined in Appendix. The extremum
condition
\begin{equation}\label{stablepoint1}
	F_0V_1-2V_0F_1=0
\end{equation}
follows from \(V_E'(0)=0\).
The absence of a ghost requires
\begin{equation}\label{Kpos1}
	K_E(0)=
	\frac{K_0}{F_0}
	+\frac{3}{2}\left(\frac{F_1}{F_0}\right)^2
	>0.
\end{equation}
Finally, the non-tachyonic condition is
\begin{equation}\label{Vpppos1}
	V_E''(0)=
	\frac{V_2}{F_0^2}
	-\frac{4V_1F_1}{F_0^3}
	-\frac{2V_0F_2}{F_0^3}
	+\frac{6V_0F_1^2}{F_0^4}
	>0.
\end{equation}
Here \(F_i\), \(V_i\), and \(K_0\) are given explicitly as integrals
over the extra-dimensional coordinate \(u\) in Appendix.
In particular, these quantities depend on the full metric profile
\(r(u;p)\) and therefore are the functions of parameters \(p\).

The three conditions --- the extremum condition~\eqref{stablepoint1}, the absence of ghosts~\eqref{Kpos1}, and the absence of tachyons~\eqref{Vpppos1} --- are necessary for the existence of a stable, ghost-free solution of the classical equations. We do not impose the vanishing of the effective potential at the extremum, \(V_E(\beta=0)=0\), which is sometimes required to ensure a vanishing cosmological constant: in a realistic setting the cosmological constant receives contributions from quantum corrections and other fields, and its smallness is a separate fine-tuning problem that lies outside the scope of this analysis. Accordingly, the only independent classical constraint on the auxiliary parameters is the extremum condition~\eqref{extr}.

\subsection{Maximally symmetric extra метрис}

The search for stable metrics consists of selecting those parameters \(p_i\) (see \eqref{param}) for which the metric \(r(u;p)\) satisfies the above conditions.
A particular metric satisfying the stability conditions has been found in \cite{Fabris:2019ecx}, and we only reproduce the result here. As shown there, maximally symmetric extra dimensions are promissing in this sense. In terms of the present paper, the initial parameters \eqref{param} are as follows:
\begin{equation}\label{sphere-ex}
	p_1=r(0)=0,\quad p_2 = r'(0)=1,\quad p_3=R'(0)=0, \quad R(0)=n(n-1)/r_{\rm sph}^2.
\end{equation}
Here the first equality places the origin at the pole of the sphere, while the second is the standard regularity condition that ensures the absence of a conical singularity. The last two equalities reflect the constant curvature of the unit \(n\)-sphere, which is uniform along \(u\).
In this case the sphere radius reads
\begin{equation}
	r_{\rm sph}=\sqrt{\bigl|2n(n-1)a+2(n-1)c_1+4c_2\bigr|}.
\end{equation}
Maximal symmetry of the solution imposes considerable limitations on the Lagrangian parameters. More precisely, the solution exists provided that the following inequalities
\begin{equation}\label{cond23}
	c_1+\dfrac{2c_2}{n-1}<0,\qquad na+c_1+\dfrac{2c_2}{n-1}<0
\end{equation}
hold.
We conclude that at least one stable solution in the form of an \(n\)-dimensional sphere exists within the framework of the model~\eqref{S}.

The physical intuition suggests that other stable solutions with an inhomogeneous radius \(r(u)\) could exist. In the next Section, the general stability argument discussed in Subsection~\ref{StabInhomo} applies to an arbitrary stable metric.

\section{Continuum of stable configurations}

The field equations governing the extra-dimensional geometry admit a family of solutions labeled by auxiliary initial data. The number of independent auxiliary parameters depends on the concrete formulation of the problem: it is determined by the number of unknown functions that appear in the chosen metric ansatz and by the possible presence of additional matter fields. Once the first integral is taken into account, this number is reduced by one. In the present paper  the three independent auxiliary parameters are chosen in \eqref{param}.
 
As shown below, whenever at least three such independent parameters remain, the single condition \eqref{stablepoint1} generically leaves a continuous two-parameter family of solutions.
Indeed, stability of a given inhomogeneous configuration \(r(u;\tilde p)\) is characterized by the conditions
\begin{equation}\label{posit}
	K_E\left(0;r(u,\tilde p)\right)>0,\qquad
	V_E''\left(0;r(u,\tilde p)\right)>0,
\end{equation}
\begin{equation}\label{stablepoint}
	F_1(\tilde p)\equiv V_E'(\beta=0;r(u,\tilde p))=0.
\end{equation}
Since the inequalities \(K_E>0\) and \(V_E''>0\) strictly hold at the point \(\tilde{p}\), they remain valid in a sufficiently small neighborhood of this point. Thus, the extremum condition~\eqref{stablepoint} is the only
independent equality constraint on the three auxiliary parameters
\(p_1,p_2,p_3\).

The equation \(F_1(p)=0\) defines, in a neighbourhood of
\(\tilde{p}\), a two-dimensional manifold in the three-dimensional
parameter space. Equivalently, one may choose two parameters, say
\(p_b\) and \(p_c\), and determine the third parameter
\[
p_a=p_a(p_b,p_c),\qquad \{a,b,c\}=\{1,2,3\},
\]
by solving the equation \(F_1(p)=0\).
Here it is assumed that the function
\(F_1(p)\) is smooth in a neighbourhood of
\(\tilde{p}\) and that \(\nabla F_1(\tilde{p})\neq0\).

A sufficiently small shift
\(\tilde{p}_b\to\tilde{p}_b+\delta p_b\),
\(\tilde{p}_c\to\tilde{p}_c+\delta p_c\)
then determines a corresponding shift
\(\tilde{p}_a\to\tilde{p}_a+\delta p_a\)
such that the extremum equation \eqref{extr} remains satisfied. In other words,
a sufficiently small variation of a stable metric
\(r(u,\tilde{p})\) within this manifold yields another stable metric
\(r(u,\tilde{p}+\delta p)\).

Since the remaining stability conditions~\eqref{posit} are strict
inequalities, their positivity persists in a sufficiently small
neighbourhood of \(\tilde{p}\). Hence, the stable configurations form
a continuous two-parameter family in a neighbourhood of \(\tilde{p}\). The set of stable metrics
is locally two-dimensional and therefore constitutes a continuum.

We conclude that radion stability does not characterize an isolated
configuration; rather, it occurs for a continuum of configurations.

More generally, for \(N\geq 2\) independent auxiliary parameters subject to the single condition \(F_1=0\), the same argument yields a locally \((N-1)\)-dimensional family of stable configurations, provided that \(\nabla F_1\) does not vanish at the stable point. For example, \(N=5\) if we also treat the warp factor \(\gamma(u)\) as an independent metric function, which adds two further initial data \(p_5=\gamma(0)\) and \(p_6=\gamma'(0)\).

A transition between two distinct static solutions \(r(u;p)\) of the landscape cannot be induced by a small perturbation localized in \(x\), because such a perturbation vanishes in the asymptotic region \(|x|\to\infty\) and therefore cannot change the asymptotic data that distinguish the solutions. Possible transitions due to non-local or non-perturbative effects, such as tunneling, are beyond the scope of the present analysis.

\section{Remark on a topological transition}

The continuum family discussed above lies within a single topological
branch of the solution space, characterized by two regular zeros of
\(r(u)\). These zeros define the positions of the branes by definition
and the endpoints of the integration interval in the extra coordinate
\(u\). Two-brane models~\cite{Tanaka:2000er} can serve as a source of
extremely massive particles localized on an otherwise empty brane
~\cite{Rubin:2025xoh} and may help explain high-energy cosmic rays.One-brane configurations have also been extensively studied in the literature, dating back four decades \cite{Akama:1982jy,Rubakov:1983bb}.
We note that the solution space may contain distinct topological
branches corresponding to one-brane and two-brane geometries for the
same action~\eqref{S} and the same physical parameters
\(a,b,c,c_1,c_2\). A sufficiently large variation of the auxiliary
parameters may change the number of regular zeros of \(r(u)\), thereby
altering the global structure of the extra-dimensional space. Such a
change suggests a transition between geometrically distinct branches.

\begin{figure}[ht!]
	\centering
	\includegraphics[width=0.4\textwidth]{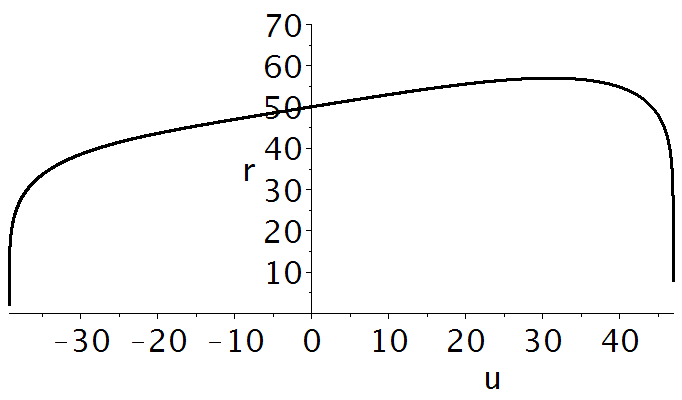}\quad
	\includegraphics[width=0.4\textwidth]{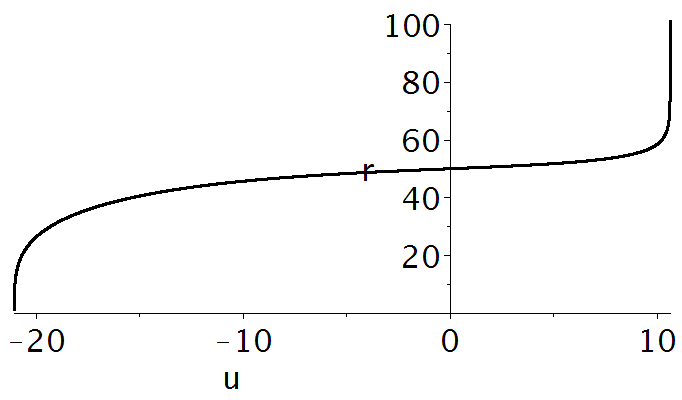}
	\caption{\small
		Plots of the extra-dimensional radius \(r(u)\) as a function of \(u\)
		for \(R'(0)=-10^{-5}\) (left panel, two-brane metric)
		and \(R'(0)=-10^{-3}\) (right panel, one-brane metric),
		with \(n=3\), \(f(R)=300R^2+R+0.002\) and \(c_1=c_2=0\).
		The remaining auxiliary conditions are fixed at
		\(r(0)=50, r'(0)=0.3,\gamma(0)=0, \gamma'(0)=0\).
		In the left panel \(r(u)\) develops two regular zeros
		(at the left and right ends of the integration interval),
		whereas in the right panel the second zero is replaced
		by a divergence of \(r(u)\), leaving a single regular zero.
	}	\label{Transit}
\end{figure}

A numerical illustration is provided in Fig.~\ref{Transit}. The left
panel shows a typical metric of the type considered in this work, while
the right panel shows the result of a moderate variation of a single
auxiliary parameter, namely the derivative of the Ricci scalar,
\[
p_3\equiv R'(0).
\]
The left panel of Fig.~\ref{Transit} shows a two-brane configuration,
while the right panel illustrates a configuration obtained after
varying \(p_3\). The two configurations belong to different branches
of the solution space, as indicated by the different numbers of regular
zeros of \(r(u)\).

The numerical examples shown here are not claimed to satisfy the
stability conditions \eqref{extr}, \eqref{Kpos}, and \eqref{Vpppos}.
They are presented only to illustrate that distinct branches of
solutions can occur for the same action and the same physical
parameters. In particular, they provide a numerical indication that
the one-brane and two-brane branches may be connected through a
variation of an auxiliary parameter.

We conjecture that each of these branches forms a continuous family
in the sense discussed above. A detailed analysis of the transition,
including the behaviour of the stability conditions
\eqref{extr}, \eqref{Kpos}, and \eqref{Vpppos} near a possible
bifurcation point, is beyond the scope of the present work.

\section{Conclusion}

We have investigated a class of higher-dimensional configurations in which the geometry of the extra dimensions is determined by the higher-dimensional field equations together with auxiliary conditions. Within this framework we derived the conditions for radion stability. These conditions are obtained here for quadratic-curvature gravity with a non-homogeneous internal metric---a case not considered previously; in contrast to the homogeneous one, they involve integrals over the extra coordinate and therefore depend on the full profile \(r(u)\), not only on its value at a single point.

Our central result is that such stable metrics are not isolated: a typical stable configuration generates a continuous two-parameter family of stable metrics in its immediate neighbourhood under sufficiently small variations of the auxiliary parameters. This follows from the fact that, among the three conditions above, only the extremum condition~\eqref{extr} is an equality; the ghost-freedom and no-tachyon conditions are strict inequalities (open conditions) and therefore do not reduce the local dimension of the solution set.

Being a solution of the system of differential equations~\eqref{EE}, the metric function \(r(u)\) and the Ricci scalar \(R(u)\) depend on the three initial conditions \(p=(p_1,p_2,p_3)\). The extremum condition~\eqref{extr} determines one of these parameters in terms of the other two, which remain arbitrary.

Stability therefore does not select a unique extra-dimensional geometry, and the resulting continuous geometrical landscape induces a corresponding continuum of effective four-dimensional parameters. This complements studies of modulus stabilization in fixed lower-dimensional geometries (e.g.\ RS-I; see the review~\cite{Bhattacharyya:2025xws} and references therein), which establish the existence of a discrete stabilized minimum but do not address whether stable solutions are isolated or continuous.

We further argued that this landscape may contain distinct topological sectors, illustrated by a possible transition between one-brane and two-brane configurations associated with a change in the number of regular zeros of \(r(u)\). Since the metrics involved differ in their asymptotic behaviour, such transitions cannot be induced by local perturbations; they would represent genuinely distinct branches of solutions rather than small fluctuations within a single branch. We emphasize that the configurations shown there are not claimed to satisfy the stability conditions; they serve only to illustrate the global structure of the solution space within which the stable continuum is embedded.

Several questions remain open. A complete spectral analysis including the massive modes present in theories with quadratic curvature invariants would go beyond the radion sector considered here. The behaviour of the continuum  near arbitrary stable solution, and the stability conditions at the bifurcation point of the topological transition, also deserve a dedicated study. Finally, combining the present stability analysis with our earlier results on the variation of observable four-dimensional parameters across the landscape~\cite{Petriakova:2023klf, Bronnikov:2023lej} would clarify the phenomenological implications of the continuous landscape. We hope to address these questions in future work.

\section*{Acknowledgment}
The work was funded by the Ministry of Science and Higher Education of the Russian Federation,
Project ”Studying physical phenomena in the micro- and macro-world to develop future technologies”
FSWU-2026-0010.

\section*{Appendix}

For later use, it is convenient to introduce the following quantities,
all of which are determined by the extra-dimensional geometry \(r(u;p)\).
We denote
\[
\mathcal{R}(u)\equiv \widetilde R_n(u)
\]
and define
\begin{align}
	F_0 &\equiv F(0)
	= v_{n-1}\int du\,r^{n-1}f_R ,
	\label{F0}
	\\
	F_1 &\equiv F'(0)
	= v_{n-1}\int du\,r^{n-1}
	\left(nf_R-2\mathcal{R}f_{RR}\right),
	\label{F1}
	\\
	F_2 &\equiv F''(0)
	= v_{n-1}\int du\,r^{n-1}
	\left[n^2f_R
	-4(n-1)\mathcal{R}f_{RR}
	+4\mathcal{R}^2f_{RRR}\right],
	\label{F2}
\end{align}
where \(f_R,f_{RR},f_{RRR}\) are evaluated at
\(R=\mathcal{R}(u)\). Similarly, defining
\[
Q(u)\equiv c_1Q_1(u)+c_2Q_2(u),
\]
we have
\begin{align}
	V_0 &\equiv V(0)
	= -\frac{v_{n-1}}{2}
	\int du\,r^{n-1}\left[f+Q\right],
	\label{V0}
	\\
	V_1 &\equiv V'(0)
	= -\frac{v_{n-1}}{2}
	\int du\,r^{n-1}
	\left[nf-2\mathcal{R}f_R+(n-4)Q\right],
	\label{V1}
	\\
	V_2 &\equiv V''(0)
	= -\frac{v_{n-1}}{2}
	\int du\,r^{n-1}
	\left[n^2f-4(n-1)\mathcal{R}f_R
	+4\mathcal{R}^2f_{RR}+(n-4)^2Q\right].
	\label{V2}
\end{align}

also
\begin{equation}
	K_0 \equiv K(0)
	=
	v_{n-1}\int du\,r^{n-1}
	\left[
	-n(n-1)f_R
	+4n\mathcal{R}f_{RR}
	+4(c_1+c_2)\mathcal{R}
	\right].
\end{equation}


\bibliographystyle{unsrtnat} 
\bibliography{Ru-Article_7}

\end{document}